\documentclass{article}

\usepackage{arxiv}

\usepackage{inputenc} 
\usepackage[T1]{fontenc}    
\usepackage{hyperref}       
\usepackage{url}            
\usepackage{booktabs}       
\usepackage{amsfonts}       
\usepackage{nicefrac}       
\usepackage{microtype}      
\usepackage{lipsum}		
\usepackage{graphicx}
\usepackage[numbers,sort&compress]{natbib}
\usepackage{doi}
\usepackage{amsmath}

\title{The Degree of Quantum Contextuality in terms of Concurrence for the KCBS Scenario}

\author{
{\hspace{1mm}Fırat Diker}
\\
	Faculty of Engineering and Natural Sciences\\
	Sabanci University\\
	Istanbul, Turkey 34956 \\
	\texttt{firatdiker@sabanciuniv.edu} \\
	\And
	{\hspace{1mm}Zafer Gedik} \\
	Faculty of Engineering and Natural Sciences\\
	Sabanci University\\
	Istanbul, Turkey 34956 \\
	\texttt{gedik@sabanciuniv.edu} \\
}

\renewcommand{\shorttitle}{The Degree of Quantum Contextuality in terms of Concurrence}

\date{}

\begin{document}
\maketitle

\begin{abstract}
	Quantum contextuality is the key concept which explains the fact that the result of a measurement is not independent of the context in which it is found. It is observed to be an intrinsic feature, i.e., neither entanglement nor spatial separation is required. In this work, we revisit the previous studies which state that entanglement is an intrinsic property called self-entanglement. Using this fact, we explicitly show the correlation between quantum contextuality and concurrence in the KCBS scenario, which is the simplest approach to observing contextuality. We also derive the equation for the maximal violation of the KCBS scenario for a given concurrence and find the linear relation between them. Using this relation, we also show how the maximal violations of the KCBS and CHSH inequalities are related for an arbitrary entanglement. Moreover, we discuss the special cases of maximal entanglement, and non-entanglement for which we have found a lower local bound.
\end{abstract}


\section{Introduction}\label{sec1}

Quantum non-locality and contextuality are unique and interesting properties of quantum systems and have attracted the scientific community's attention since the pioneering works by Bell, Gleason, Kochen, and Specker \cite{Gleason,1_Bell,2,3}.  A solid understanding of these concepts plays a vital role in Quantum Theory (QT). Discoveries of these characteristics are milestones separating quantum physics from the classical world; because the context in which a measurement is performed does affect the outcome in quantum physics whereas that is not the case for a classical measurement. They still baffle researchers who have put much effort into this endeavor. Discussions have arisen concerning the completeness of QT. Therefore, the possibility of hidden variables has been suggested, which have later been denied by various works \cite{1_Bell,2,3,4_Bell,5,6}. The most basic examples of classical inequalities are the Klyachko-Can-Binicio\u{g}lu-Shumovsky (KCBS) inequality, and the Bell inequality \cite{1_Bell,KCBS1,KCBS2,KCBS3,KCBS4,reviewer}. These measurement scenarios concern the fundamental aspects of QT. We will mainly discuss the first scenario where an observer performs measurements on qutrit states (three-level quantum systems) to observe contextuality. The result of this work and later experimental realizations have shown us that some qutrit states are intrinsically contextual and compatible with QT without the need for entanglement \cite{8,9}. What is more, it has been shown that every qutrit state can exhibit contextual behavior if the number of measurements is increased \cite{10,11}.

The outline of this paper is as follows: In this section, we will give a brief introduction to the KCBS inequality and discuss its violation. In Section $II$, we will introduce Majorana Stellar Representation (MSR) \cite{12,13,15} of two qubits and use it to define effective qutrits. MSR is beneficial for the geometrical construction of the mutually unbiased bases (MUBs), and  symmetric informationally complete positive operator valued measures (SIC-POVMs) of a spin-1 system \cite{Ara}. The qutrits of MSR will be used to calculate the expectation value of the KCBS operator. Furthermore, we will calculate the concurrence of two qubits and then derive the mathematical relation between the average value of the KCBS operator \cite{KCBS1,KCBS2,KCBS3,KCBS4} and concurrence \cite{16}. From this relation, we will derive the formula of the maximal contextuality for a given concurrence. We will make use of this formula to find the relation between the maximal violations of both the KCBS and CHSH inequalities. For the non-entanglement case, we will find a new lower limit. We will also discuss how contextuality and entanglement are connected in the KCBS scenario. In Section $III$, we will conclude our results and categorize qutrits based on their violations of the non-contextuality and local bounds. 

Bell has proposed the locality inequality to test if local restrictions are valid and shown that some quantum systems exhibit non-locality \cite{1_Bell}. The other interesting feature is that any entangled state violates this inequality \cite{17,conc-bell1, conc-bell2}. Yet, this is not always the case since there are some Bell inequalities that are not violated by entangled states \cite{not-violated1,not-violated2,not-violated3,not-violated4}. This is the direct result of non-locality, which is unique to QT. Note that Bell measurements are performed on bipartite states in a $4$-dim Hilbert space ($d^2 \otimes d^2$). The next step was to show whether one can prove it in a Hilbert space with less than $4$ dimensions. This has been achieved by Kochen, and Specker \cite{2,3}. One can observe a non-classical behavior of a quantum state in a $3$-dim Hilbert space, which is called contextuality and is unique to QT. This discovery took us a step further to understand the quantum paradigm better. 

After years, the most simplified version of a contextuality test has been found, which forms the basis of our work. This is called the KCBS test including five measurements \cite{KCBS1,KCBS2,KCBS3,KCBS4}. It has been shown in \cite{KCBS1} at first, and then discussed extensively in \cite{KCBS4}. This test consists of a state-dependent non-contextuality inequality,
\begin{equation}
\langle A_1 A_2 \rangle + \langle A_2 A_3 \rangle + \langle A_3 A_4 \rangle + \langle A_4 A_5 \rangle + \langle A_5 A_1 \rangle \geq -3
\label{eq1}
\end{equation}
where $A_i = 2 S_i^2 - 1$ and $S_i$ are the usual spin-$1$ operators. One may find five vectors along which spin measurements are performed in the physical Euclidean space $\mathbb{E}^3$ so that $A_i$ and $A_{i+1}$ can be measured together. These vectors form a pentagram, and the geometric form of the inequality is named after this. Each term is the expected value of measurements of two orthogonal observables. The left-hand side gives results exceeding the classical lower limit; in other words, qutrit states are intrinsically contextual without being entangled. One may observe a maximal violation of the above inequality for the neutrally polarized spin state $\vert 0 \rangle$. The quantum lower limit is close to $-4$ ($\cong -3,94$). Please refer to Appendix \ref{secA1} for details.

\section{The expectation value of the KCBS operator in terms of Concurrence}\label{sec2}

The violation of the KCBS inequality shows the contextual property for qutrits, and it is defined as an intrinsic feature; in other words, one does not need the spatial separation of two or more particles. Naturally, a question arises whether one may quantify entanglement as an intrinsic property. It has already been shown that a qutrit can have embedded entanglement, i.e., self-entanglement \cite{18}. The authors have used the symmetric group of two-qubit states to define effective qutrits and found the concurrence inequality as a measure of self-entanglement for qutrits \cite{KCBS3}. 

\begin{figure}[t!]
\centering
\includegraphics[width=0.6\linewidth]{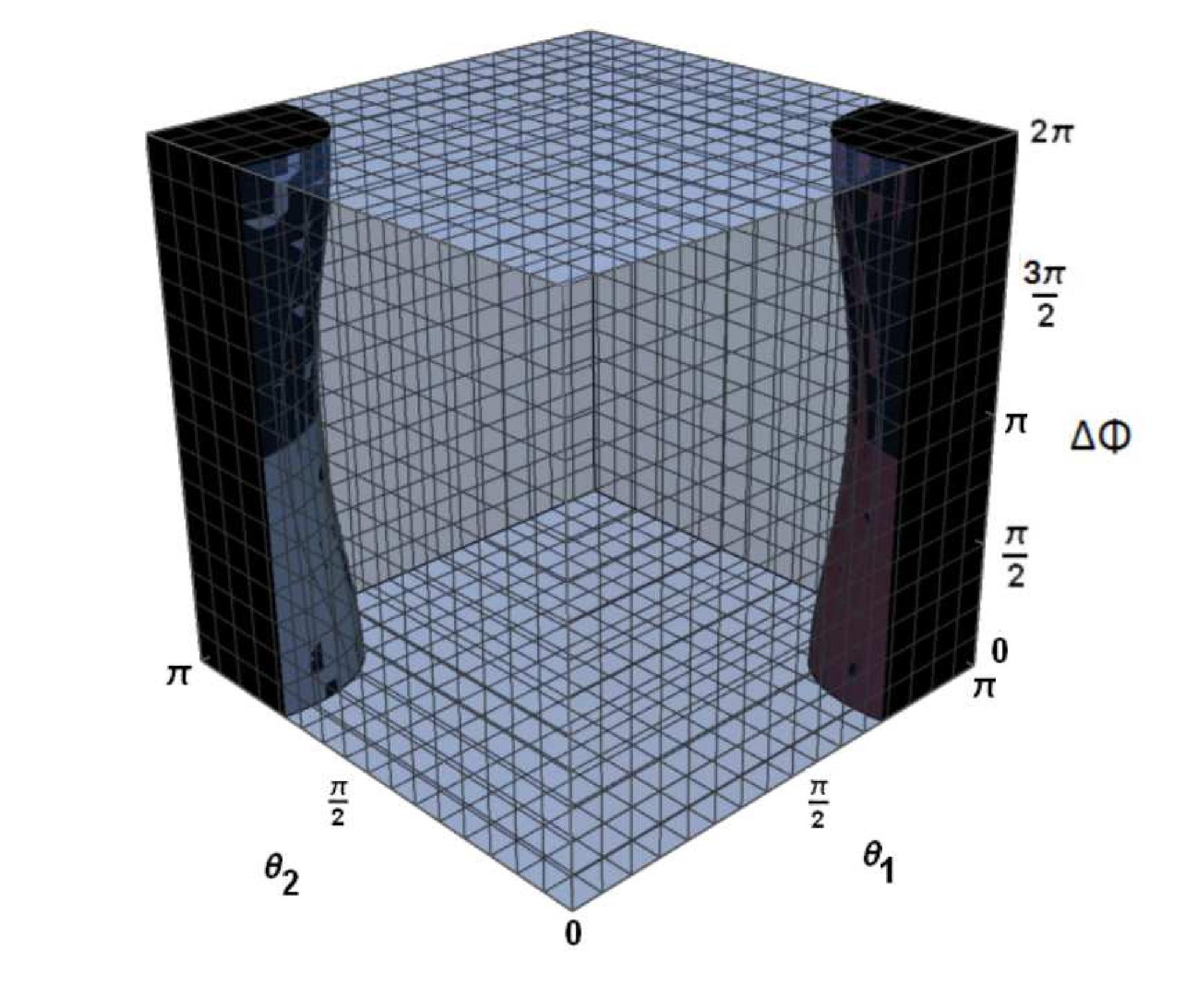}
\caption{(color online) This is the illustration of (non-)contextuality regions of the $S$ function in terms of $\theta_i$ and $\Delta \phi$ ($i=1$,$2$). The black part is the contextuality region ($S<-3$); the blue translucent part is the non-contextuality region ($S>-3$). The maximum violation of the KCBS inequality is found for $\theta_1 = 0$, $\theta_2=\pi$ and $\theta_1 = \pi$, $\theta_2=0$ (independent of $\Delta \phi$). These correspond to the two edges of the rectangular cuboid in the black region.}
\label{fig1}
\end{figure}

We will also make use of MSR for the symmetric group to define effective qutrits. The set of symmetric two-qubit states corresponding to effective qutrits that violate the KCBS inequality have already been investigated \cite{set}. We do the same here by using the different representation. The advantage of MSR is that one can work with the whole group of qutrits by using this representation. Please refer to Appendix \ref{secA2} for details on the symmetric states and their concurrence. MSR of any qubit corresponds to a vector from origin to a point on Bloch sphere, making easier to visualize a qubit. Here, we need to use the symmetric subgroup of two-qubit states. The symmetric expression of two majorana particles is as follows: $\vert\psi\rangle = \frac{1}{N} ( \vert m \rangle \vert n \rangle + \vert n \rangle \vert m \rangle)$ where 
$\vert m \rangle, \vert n \rangle= (
 \cos \frac{\theta_{1,2} }{2} \  \
 \sin \frac{\theta_{1,2} }{2} e^{i {\phi}_{1,2}} )$.
As long as two-qubit states belong to the symmetric subgroup, we may define them as effective qutrits \cite{18}. Next step is to take the average of the KCBS operator for effective qutrits, which gives
\begin{equation}
    \langle \psi \vert \mathbb{S}_{KCBS} \vert \psi \rangle = \frac{4 \left(3 \sqrt{5}-5\right) (\cos \text{$\theta_1 $} \cos \text{$\theta_2 $}+1)}{\sin \text{$\theta_1 $} \sin \text{$\theta_2 $} \cos (\text{$\phi_1 $}-\text{$\phi_2 $})+\cos \text{$\theta_1 $} \cos \text{$\theta_2 $}+3} + (5-4 \sqrt{5}).
    \label{eq4}
\end{equation}
That is a function of $\theta_1$, $\theta_2$ and $\Delta \phi$ ($\Delta \phi = \text{$\phi_1 $}-\text{$\phi_2 $}$). For details on how one can reach Equation \ref{eq4} (the $S$ function), please read Appendix \ref{secA3}. In Figure \ref{fig1}, by using the $S$ function, we illustrate the distinction between contextual and non-contextual states in a $3$-dim plot. The numeric part of the $S$ function in Equation \ref{eq4} equals to the lowest value possible ($\cong -3,94$), i.e.,  it corresponds to the maximal violation of the non-contextuality inequality. We obtain the minimum value of the $S$ function when $\cos \text{$\theta_1 $} \cos \text{$\theta_2 $} = -1$($\theta_1 = \pi$ and $\theta_2 = 0$, or vice versa). In a sense, the numeric part acts as a reference point to determine the degree of contextuality (recall that $-3$ is the classical lower limit). The importance of the $S$ function given in Equation \ref{eq4} comes from the fact that it shows if a qutrit state exhibits contextuality, and if yes, it informs us on the degree of contextuality.

Now we will calculate the concurrence for symmetric two-qubit states obeying the general form in \cite{16}. The concurrence can be expressed for MSR of two particles as below:
\begin{equation}
    C(\theta_1, \theta_2, \Delta \phi) = \frac{1 - \sin \text{$\theta_1 $} \sin \text{$\theta_2 $} \cos \text{$\Delta \phi $} - \cos \text{$\theta_1 $} \cos \text{$\theta_2 $} }{3 + \sin \text{$\theta_1 $} \sin \text{$\theta_2 $} \cos \text{$\Delta \phi $} + \cos \text{$\theta_1 $} \cos \text{$\theta_2 $}}.
    \label{eqC}
\end{equation}
This can be recast into the following form:
\begin{equation}
C[f(\theta_1, \theta_2, \Delta \phi)]= \frac{1 - f(\theta_1, \theta_2, \Delta \phi) }{3+f(\theta_1, \theta_2, \Delta \phi)}
\end{equation}
where 
\begin{equation}
f(\theta_1, \theta_2, \Delta \phi) = \sin \text{$\theta_1 $} \sin \text{$\theta_2 $} \cos \Delta \phi + \cos \text{$\theta_1 $} \cos \text{$\theta_2 $}.
\label{eq7}
\end{equation}
Equation \ref{eq7} is equal to $\cos {2 \theta_{mn}} $ where $\theta_{mn}$ is the angle between state vectors, $\vert m \rangle$ and $\vert n \rangle$, which should not be confused with $\theta_{1(2)}$. Both Equations \ref{eq4} and \ref{eqC} can be written in terms of $\theta_{mn}$. This tells us that the degree of violation in the KCBS scenario and the concurrence of the corresponding state depends on relative angle between these state vectors, $\vert m \rangle$ and $\vert n \rangle$. One may easily define Equation \ref{eq7} in terms of concurrence,
\begin{equation}
f(\theta_1, \theta_2, \Delta \phi) = \frac{1-3 C(\theta_1, \theta_2, \Delta \phi)}{1 + C(\theta_1, \theta_2, \Delta \phi)}
\end{equation}
which we plug in the $S$ function given in Equation \ref{eq4}, and get
\begin{equation}
S(\text{$\theta_1 $},\text{$\theta_2 $}, \Delta \phi) = \left(3 \sqrt{5}-5\right) (C(\theta_1, \theta_2, \Delta \phi)+1) (\cos \text{$\theta_1 $} \cos \text{$\theta_2 $} + 1)+5-4 \sqrt{5}.
\label{eq9}
\end{equation}
This is the direct relation between the degree of entanglement and contextuality. Since concurrence is a function of $\theta$'s and $\Delta \phi$, other cosine terms are not independent of concurrence. 
As one can see, we still have the cosine terms making Equation \ref{eq9} more complicated. We would prefer it to be simpler, but that is not the case. 

Our next step is to find the lower bound of the $S$ function for a given concurrence. For this, Equation \ref{eq7} must be taken constant so that concurrence does not change. We have two variables($\theta_{i=1,2}$) while $C$ is constant. To keep $C$ constant while changing them, we require the following condition:
\begin{equation}
\cos (\Delta \phi )=\frac{\frac{1-3 C}{1+C}-\cos \theta_1 \cos \theta_2}{\sin \theta_1 \sin \theta_2},
\label{deltaphi}
\end{equation}
and $-1\leq \cos (\Delta \phi ) \leq 1$. The coefficient of the part including cosine terms in Equation \ref{eq9} is positive. We need to adjust $\theta$ values such that the cosine part is as small as possible because we aim at finding the minimum value of the $S$ function when $C$ is known. We need to take the limit for $\theta_1 \xrightarrow[]{} 0 $ and $\theta_2 \xrightarrow[]{} \pi $, or vice versa due to symmetry. We see that $\lim_{\substack{\theta_1\to0 \\ \theta_2\to\pi}} \cos (\Delta \phi ) = \pm \infty$, which cannot be true due to trigonometric restrictions. We need to find the $\theta$ values which give us the upper and lower bounds for Equation \ref{deltaphi} ($-1\leq \cos (\Delta \phi ) \leq 1$). When we solve for $\cos (\Delta \phi ) = \pm 1$, we obtain $\cos ^{-1}\left(\frac{1-3 C}{1+C}\right)=\theta_1-\theta_2$ and $\cos ^{-1}\left(\frac{1-3 C}{1+C}\right)=\theta_1 + \theta_2$. Using these relations, we may express $\theta_1$ in terms of $\theta_2$ and $C$. Recall that the average value of the KCBS operator, $S(\text{$\theta_1 $},\text{$\theta_2 $}, \Delta \phi)$ has three variables. Since $C$ is taken constant and $\theta$s can be written in terms of each other, we reduce the number of variables from three to one. Therefore, the $S$ function may be written as $S(\theta_i)$ ($i=1$ or $2$). Notice that we look into the special case for $S(\theta_i)_{min}$ provided that $\cos (\Delta \phi ) = \pm 1$. Then, we solve for $d S (\theta_2) / d \theta_2 = 0$ and obtain the following solution set: $\theta_2 \in \{ {\frac{1}{2} \left(\pi + \cos ^{-1}\left(\frac{1-3 C}{1+C}\right) \right), \frac{1}{2} \left(\pi -\cos ^{-1}\left(\frac{1-3 C}{1+C}\right)\right)} \}$. Finally, we get the following for both $\theta_2$ values:
\begin{equation}
S^{min}_{KCBS} = \left(5-3 \sqrt{5}\right) C - \sqrt{5}.
\label{SminC}
\end{equation}
This gives the minimum value of the $S$ function for a given concurrence. Equation \ref{SminC} shows us how much a qutrit state can violate the KCBS inequality when the degree of entanglement is constant. There is a linear relation between quantum contextuality and entanglement. We know that $0 \leq C \leq 1$, and one may easily check for $C = 0$ and $C = 1$, i.e., the non-entanglement and maximal entanglement cases, respectively. For $C=0$, $S^{min}_{KCBS} = - \sqrt{5}$ which gives the lower limit for local (separable) states. This is the classical lower bound under local restrictions and is a higher limit than the previous one, imposing a stricter separation between classicality and quantumness. Below this new limit, we observe only entangled states. When two particles are maximally entangled ($C=1$), $S^{min}_{KCBS} = 5-4 \sqrt{5} $ which is the quantum lower limit for the KCBS scenario. For the non-contextuality bound $S_{KCBS}=-3$, we find that $C \cong 0.447$. That is quite high, and qubits must be well entangled for observers to see quantum contextuality. This gives insights into the connection between contextuality and entanglement. 

One may also calculate the maximum value of the $S$ function when concurrence is constant. As mentioned earlier, the part which includes the cosine terms in the $S$ function has a positive coefficient. We need $\theta$ values to make that part as high as possible since we try to find the maximum value. The part with the cosine terms is at maximum for $\theta_{1(2)}=0$, or $\theta_{1(2)}=\pi$. Due to trigonometric restrictions imposed by Equation \ref{deltaphi}, we cannot set these values. In accordance with the restriction ($-1\leq \cos (\Delta \phi ) \leq 1$), we obtain the same solution set of $\theta$ values: $\cos ^{-1}\left(\frac{1-3 C}{1+C}\right)=\theta_1-\theta_2$ and $\cos ^{-1}\left(\frac{1-3 C}{1+C}\right)=\theta_1 + \theta_2$. However, we will now look into the subset of $\theta$ values for the maximal $S$ function. We may write the $S$ function as $S(\theta_2)$ and solve for $d S (\theta_2) / d \theta_2 = 0$. This gives us the following subset for the maximal $S$: $\theta_2 \in \{\frac{1}{2} \cos ^{-1}\left(\frac{1-3 C}{1+C}\right),  -\frac{1}{2} \cos ^{-1}\left(\frac{1-3C}{1+C}\right)  \} $. When we plug these values in the $S$ function, we get the following: $S_{\text{max}}(\theta_2) = -0.53$. This result tells us that the measurements may always yield outcomes within the classical range, which does not depend on whether a qutrit state is entangled or not. There is not such a state always exhibiting contextuality due to its high degree of entanglement. Moreover, the maximal value does not depend on concurrence, so the higher limit is the same for all qutrit states.

Similarly, the maximal violation of the CHSH inequality for a known concurrence has been given by the following formula, $\beta = 2 \sqrt{1+C^2}$ where $\beta$ is the average value of the CHSH operator \cite{conc-bell1,conc-bell2}. One also finds that the maximal violations of both inequalities obey the following proportionality: $S \propto - \sqrt{\beta ^2 - 4}$. We may interpret it as the direct relation between contextuality and non-locality for extreme cases.

\section{Conclusion}\label{sec3}

We have discussed quantum contextuality for the symmetric subgroup of two-qubit states expressed in MSR and shown their degree of contextuality in the KCBS scenario. Later, we have found a connection between contextuality and concurrence, which has been shown in Equation \ref{eq9}. Using this relation, we have found the maximal quantum contextuality in terms of a given concurrence (the measure of entanglement). We have seen that there is a linear relation between them. We have also looked into the maximal and no entanglement cases that give the quantum and the classical lower limits, respectively. These limits tell us that the violation of the non-contextuality inequality requires quantum systems to be not just contextual but also non-local in this measurement scenario. This has been discussed in \cite{KCBS2, KCBS4} and it is already known that there is a correlation between the degree of entanglement and the violation of the KCBS inequality. What we have achieved is that this correlation has been shown explicitly. We can divide the group of qutrit states into three categories: (1) The qutrits which do not violate the KCBS inequality; (2) The states which violate the KCBS inequality to a certain degree (not maximally); (3) The maximal violation; in other words, those that yield the lowest result possible (the quantum lower limit). We may add subcategories dividing qutrits into: (1) contextual and non-local group for $S<-3$; (2) Those that do not violate the KCBS inequality but exhibits non-locality ($ -3 \leq S < -\sqrt{5} $); (3) Those that do not violate the KCBS inequality and exhibit locality ($ -\sqrt{5} \leq S $). These categories are determined by the relation between the violation of the KCBS inequality and concurrence. The same categorization can be provided using the results based on the relation between the KCBS and CHSH inequalities \cite{set}. Categorization of qutrit states based on their quantumness is of the essence for further research, and each step taken in this process allows us to understand the quantum paradigm better. Moreover, the KCBS scenario was a convenient choice for this purpose since it is the most basic example of a non-contextuality inequality. As mentioned earlier, one may observe contextuality for every qutrit state when the number of measurements is increased; in other words, without regarding the degree of self-entanglement (or concurrence), qutrit states exhibit contextual behavior \cite{10,11}. This is also important because dependency on entanglement is observed by reducing the number of measurements, which is what happens in our case.

\appendix

\section{The KCBS Inequality}\label{secA1}

\begin{figure}[t!]
\centering
  \includegraphics[width=0.6\linewidth]{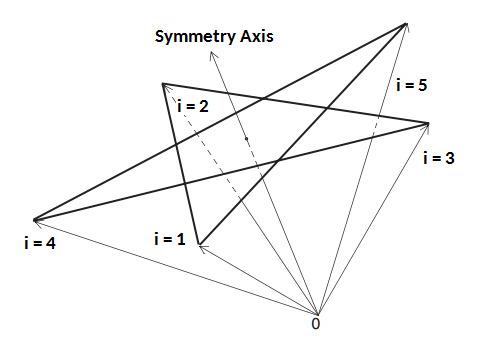}
\caption{In the KCBS scenario, there are five measurements in total, performed along five vectors ($i = 1, ... , 5$) in the physical space.}
  \label{pentagram}
\end{figure}

We will look into how the KCBS inequality is formed using the spin-1 measurements \cite{KCBS1,KCBS2,KCBS3,KCBS4}. Recall that the inequality is given as
\begin{equation}
    \langle A_1 A_2 \rangle + \langle A_2 A_3 \rangle + \langle A_3 A_4 \rangle + \langle A_4 A_5 \rangle + \langle A_5 A_1 \rangle \geq -3
\end{equation}
where 
\begin{equation}
    A_i = 2 {S_i}^2 - I.
\end{equation}
$S_i$ are the spin-1 measurements, and $I$ is the identity matrix. The sub-index $i$ is the direction of each spin measurement (five in total).The vectors along these directions form a pentagram as can be seen in Figure \ref{pentagram}. There are three eigenvalues for each spin measurement:
\begin{equation}
    S_i  \vert s \rangle = s \vert s \rangle
\end{equation}
where $s \in \{ 0, \pm 1 \}$ and the ket $\vert s \rangle $ is the corresponding eigenstate. Therefore, if one measures the new observable $A_i$ for each eigenket, we get
\begin{equation}
    A_i \vert s \rangle = (2 s^2 - 1) \vert s \rangle.
\end{equation}
Defining the new observable allows us to reduce the number of eigenvalues from three to two, which can be seen in the following:
\begin{equation}
    A_i \vert \pm 1 \rangle = \vert \pm 1 \rangle,
\end{equation}
and
\begin{equation}
    A_i \vert 0 \rangle = - \vert 0 \rangle.
\end{equation}
The directions of $S_i$ and $S_{i+1}$ are perpendicular to each other. Therefore, $A_i$ and $A_{i+1}$ are compatible and can be measured together. If one of the spin measurements yields $+1$, the other one must give $-1$ due to orthogonality. The KCBS inequality is always satisfied for classical states, yielding results above the lower limit. Each of the $\langle A_i A_{i+1} \rangle$ must be $-1$ if one wants the left-hand side of Equation A1 to be as low as possible; however, one of the pairs in each case is equal to $+1$ while the other terms are equal to $-4$ in total. This gives us the classical lower limit, $-3$. 

\section{The symmetric subgroup of two-qubit states and its concurrence}\label{secA2}

We will briefly discuss the symmetric subgroup of two-qubit states corresponding the whole group of qutrits in the $3$-dim Hilbert space. The general form of the symmetric two-qubit states is given as
\begin{equation}
    \vert \Psi \rangle = \boldsymbol{\alpha_1} \vert 00 \rangle + \boldsymbol{\beta} \bigg( \frac{\vert  01 \rangle + \vert 10 \rangle}{\sqrt{2}} \bigg) +  \boldsymbol{\alpha_2} \vert 11 \rangle
    \label{B7}
\end{equation} 
where $\boldsymbol{\alpha_i}$ ($i=1,2$) and $\boldsymbol{\beta}$ are complex probability amplitudes. The kets $\vert 0 \rangle$ and $\vert 1 \rangle$ correspond to the particles with spins $1/2$ and $-1/2$, respectively. Thus, the kets $\vert 00 \rangle$ and $\vert 11 \rangle$ have the total spin of $1$ and $-1$, respectively. It is trivial to say that the kets $\vert 01 \rangle$ and $\vert 10 \rangle$ are zero-spin states. The total spin of each ket corresponds to an eigenvalue of a spin-$1$ operator, which allows us to rewrite Equation \ref{B7} as follows:
\begin{equation}
    \vert \Psi^* \rangle = \boldsymbol{\alpha_1} \vert 1 \rangle + \boldsymbol{\beta} \vert  0 \rangle +  \boldsymbol{\alpha_2} \vert -1 \rangle
    \label{B8}
\end{equation} 
One should not confuse the spin-1 states $\vert  0 \rangle$ and $\vert  1 \rangle$ in Equation \ref{B8} with the ones corresponding to the spin-$1/2$ particles in Equation \ref{B7}. 

After defining two-qubit states as effective qutrits, one may calculate concurrence as a measure of self-entanglement for qutrits \cite{KCBS3}. For this, we apply the general concurrence formula $ \langle \Psi \vert \Tilde{\Psi} \rangle $ which gives
\begin{equation}
    C (\Psi) = 2 \vert \boldsymbol{\alpha_1} \boldsymbol{\alpha_2} - {\boldsymbol{\beta}}^2 \vert.
\end{equation}
Using symmetric two-qubit states allows us to calculate concurrence for qutrits, which is called self-entanglement, implying that entanglement is an intrinsic feature.

\section{The calculation of the S function}\label{secA3}

In this section, we will introduce the KCBS operator composed of the observables $A_i$ as shown in Equation \ref{eq1} and show how we have reached the $S$ function expressed in Equation \ref{eq4} which gives the expectation value of the KCBS operator. 

The KCBS operator is simply a diagonal $3 \times 3$ matrix with matrix elements $S_{11} = S_{33} = 2 \sqrt{5}-5$ and $S_{22} = 5-4 \sqrt{5}$. The average of the KCBS operator is found for the symmetric two-qubit states shown in MSR as in the following:
\begin{equation}
    \langle \psi \vert \mathbb{S}_{KCBS} \vert \psi \rangle = 
 {\left(
\begin{array}{ccc}
 {\psi_1}^*  \\
 {\psi_2}^*  \\
 {\psi_3}^*  \\
\end{array}
\right)}^T  
    \left(
\begin{array}{ccc}
 2 \sqrt{5}-5 & 0 & 0 \\
 0 & 5-4 \sqrt{5} & 0 \\
 0 & 0 & 2 \sqrt{5}-5 \\
\end{array}
\right) 
\left(
\begin{array}{ccc}
 \psi_1  \\
 \psi_2  \\
 \psi_3  \\
\end{array}
\right) 
\label{C10}
\end{equation}
where
\begin{equation}
    \vert \psi \rangle = \frac{1}{N} \left(
\begin{array}{c}
 \cos \left(\frac{\theta_1}{2}\right) \cos \left(\frac{\theta_2}{2}\right) \\
 \frac{1}{\sqrt{2}} (e^{i \text{$\phi_1$}} \cos \left(\frac{\text{$\theta_2$}}{2}\right) \sin \left(\frac{\text{$\theta_1 $}}{2}\right)+e^{i \text{$\phi_2 $}} \cos \left(\frac{\text{$\theta_1 $}}{2}\right) \sin \left(\frac{\text{$\theta_2 $}}{2}\right)) \\
 e^{i (\text{$\phi_1 $}+\text{$\phi_2 $})} \sin \left(\frac{\text{$\theta_1 $}}{2}\right) \sin \left(\frac{\text{$\theta_2 $}}{2}\right) \\
\end{array}
\right).
\end{equation}
This is MSR of a two-qubit state with the normalization constant
\begin{equation}
    N = \sqrt{\frac{1}{4} (\sin \text{$\theta_1 $} \sin \text{$\theta_2 $} \cos \left( \text{$\phi_1 $}-\text{$\phi_2 $} \right) + \cos \text{$\theta_1 $} \cos \text{$\theta_2 $} +3)}.
\end{equation}
From Equation \ref{C10}, we get 
\begin{equation}
\begin{split}
    \langle \psi \vert \mathbb{S}_{KCBS} \vert \psi \rangle = \frac{1}{N^2} {\left(
\begin{array}{c}
 \cos \left(\frac{\theta_1}{2}\right) \cos \left(\frac{\theta_2}{2}\right) \\
 \frac{1}{\sqrt{2}} (e^{-i \text{$\phi_1$}} \cos \left(\frac{\text{$\theta_2$}}{2}\right) \sin \left(\frac{\text{$\theta_1 $}}{2}\right)+e^{-i \text{$\phi_2 $}} \cos \left(\frac{\text{$\theta_1 $}}{2}\right) \sin \left(\frac{\text{$\theta_2 $}}{2}\right)) \\
 e^{-i (\text{$\phi_1 $}+\text{$\phi_2 $})} \sin \left(\frac{\text{$\theta_1 $}}{2}\right) \sin \left(\frac{\text{$\theta_2 $}}{2}\right) \\
\end{array}
\right)}^T  \\
. \left(
\begin{array}{c}
 \left(2 \sqrt{5}-5\right) \cos \left(\frac{\theta_1}{2}\right) \cos \left(\frac{\theta_2}{2}\right) \\
 \frac{1}{\sqrt{2}} (\left(5-4 \sqrt{5}\right) \left(e^{i \phi_1} \cos \left(\frac{\theta_ 2}{2}\right) \sin \left(\frac{\theta _1}{2}\right)+e^{i \phi_2} \cos \left(\frac{\theta_1}{2}\right) \sin \left(\frac{\theta_2}{2}\right)\right)) \\
 \left(2 \sqrt{5}-5\right) e^{i (\phi_1+\phi_2)} \sin \left(\frac{\theta_1}{2}\right) \sin \left(\frac{\theta_2}{2}\right) \\
\end{array}
\right) 
\end{split}
\end{equation}
which gives 
\begin{equation}
\begin{split}
   S(\theta_1, & \theta_2, \Delta \phi)  = \\
  & \frac{\left(5-4 \sqrt{5}\right) \sin (\theta_1) \sin (\theta_2) \cos (\phi_1 - \phi_2)+\left(8 \sqrt{5}-15\right) \cos (\theta_1) \cos (\theta_2)-5}{\sin (\theta_1) \sin (\theta_2) \cos (\phi_1-\phi_2)+\cos (\theta_1) \cos (\theta_2)+3}.
\end{split}
\end{equation}
This result can be formed into the $S$ function given in Equation \ref{eq4}.

\bibliographystyle{unsrt}







\end{document}